# Synthesizing extreme-ultraviolet vector beams in a chip

Riccardo Piccoli,[1,2,*] Marco Bardellini,[1] Stavroula Vovla,[1] Linda Oberti,[1] Kamal A. A. Abedin,[1] Anna G. Ciriolo,[3] Rebeca Martínez Vázquez,[3] Roberto Osellame,[3] Luca Poletto,[4] Fabio Frassetto,[4] Davide Faccialá,[3] Michele Devetta,[3] Caterina Vozzi,[3] and Salvatore Stagira[1]

[1]*Department of Physics, Politecnico di Milano, Milano, I-20133, Italy*
[2]*Department of Molecular Sciences and Nanosystems, Ca' Foscari University of Venice, Venice, 30172 Italy*
[3]*National Research Council (CNR), Institute for Photonics and Nanotechnologies, Milano, I-20133, Italy*
[4]*National Research Council (CNR), Institute for Photonics and Nanotechnologies, Padua, I-35131 Italy*
*\*riccardo.piccoli@unive.it*

**Abstract:** Structured light has gained significant attention in recent years, especially in the generation and application of vector beams. These beams, characterized by a spatially varying polarization state, are a powerful tool to enhance our capacity to control light-matter interactions. In this study, we demonstrate the synthesis of extreme-ultraviolet (EUV) vector beams in a chip through high-order harmonic generation (HHG). Our findings showcase the chip's ability to transfer the laser polarization state into the EUV beam despite the extended interaction length. This approach not only outperforms conventional free-space methods but also paves the way for a multitude of on-chip investigations in the realms of EUV and soft-X-ray spectroscopy.

## 1. Introduction

The polarization of light is a fundamental parameter that shapes the very nature of how light interacts with matter. Within this realm, laser beams featuring spatially variant polarization states, commonly known as vector beams [1,2], emerge as a powerful tool providing unparalleled control over both classical and quantum phenomena. From well-known applications in high-resolution microscopy [3], sub-diffraction [4] and optimal plasmonic focusing [5], particle acceleration [6] and trapping [7,8], as well as laser micro-machining [9,10], vector beams have extended their capabilities. Nowadays, they encompass three-dimensional vectorial holography [11], generation of hybrid-entangled states [12], multiple-degree-of-freedom quantum memory [13], large-capacity data storage [14], excitation of magnetic phenomena [15,16], and beyond, finding diverse and impactful applications.

The generation of vector beams in the visible and infrared domains has seen advancements through various techniques, including laser cavity methods [17], interferometric techniques [18], segmented waveplates [19], metasurfaces [20], and spatial light modulators [21]. Nowadays, liquid crystal devices with spatially varying molecular orientation [22] provide a straightforward means for achieving this. However, these techniques face limitations when it comes to manipulating polarization in the extreme-ultraviolet (EUV) domain, at photon energies ranging from tens to hundreds of electronvolts, posing a significant challenge associated with substantial losses [23]. It is worth mentioning the very recent demonstration of meta-lens operating in the EUV domain [24], which paves the way for wavefront manipulation in this range of frequencies. Nonetheless, such approach remains inherently constrained to a narrow bandwidth centered around the designated frequency.

A recent method for overcoming these issues exploits the high-order harmonic generation (HHG) process to up-convert light from the infrared-visible range to the EUV. Indeed, the polarization state of a vector beam is locally linear and thus it is directly transferred by the HHG process to the EUV beam. The preservation of polarization and phase spatial distribution in the generation of EUV vector beams through HHG has been demonstrated using gas jets or gas cells and free-space laser propagation in a loose focusing geometry [25–30].

Scaling upward the inherently low harmonic generation yield calls for strategies based on long interaction lengths, which cannot be achieved in gas jets. On the other hand, long gas cells require multi-millijoule laser systems to maintain high peak intensities over long Rayleigh ranges in favorable phase-matching conditions.

Here we propose a breakthrough in the generation of EUV vector beams based on HHG in a novel microfluidic device, hereafter called *chip* for conciseness [31,32]. Fabricated through the *Femtosecond Laser Irradiation and Chemical Etching* (FLICE) technique [33], this device allows extended interaction lengths by guiding the intense light pulses in a hollow waveguide filled with gas. Such an approach has already surpassed conventional gas jet methods by demonstrating enhanced high harmonics conversion efficiency with linearly polarized drivers [32]. The hollow-core waveguide supports the propagation of several spatial modes characterized by distinct polarization states: hybrid modes ($EH_{nm}$ – linear polarization), transverse magnetic ($TM_{0m}$ – radial polarization) modes, and transverse electric ($TE_{0m}$ – azimuthal polarization) modes [34].

In this work, we leverage the frequently overlooked multimodal nature of a hollow waveguide as a key asset. Specifically, we demonstrate the successful decomposition of a laser vector beam into a superposition of $TM_{01}$ and $TE_{01}$ modes inside the device, subsequently upconverted into the EUV domain via HHG. Our findings not only affirm the feasibility of on-chip EUV vector beam generation but also open avenues for controlling vector beam properties through the direct manipulation of the waveguide's modal characteristics. This work marks a notable advancement in the intersection of structured light research and the realm of photonic integration, showing a promising pathway toward the development of high-brightness integrated sources of EUV beams carrying complex states of light.

## 2. Results

Our investigation centers on a laser vector beam focused into a chip fed with argon gas (Fig. 1a). The hollow waveguide can sustain different spatial modes, including the $TM_{01}$ and $TE_{01}$ modes (Fig. 1b). As the laser beam propagates through the channel, it creates EUV light via HHG, resulting in an output vector beam with the same polarization state as the driver. Figure 1c presents a comprehensive illustration of the entire experimental setup. An amplified Ti:Sapphire laser system (Amplitude ARCO) emits multi-millijoule pulses of about 25 fs duration (full-width at half maximum – FWHM) at a wavelength of 800 nm and 1 kHz repetition rate. The laser beam propagates through a vortex waveplate (zero-order vortex half-wave retarders from *Thorlabs*) for polarization state modulation. This allows us to set the polarization state to, e.g., radial, depending on the vortex waveplate orientation relative to the laser beam input polarization. It is crucial to highlight that the vortex waveplate only impacts the polarization state without introducing a spatial phase, such as the helical phase associated with beams carrying orbital angular momentum. After modifying its polarization state, the vector beam is transported towards the chip by metallic mirrors, carefully placed to minimize incident angles and reduce depolarization effects. A half-waveplate (HWP) on a motorized rotation stage is placed in the path as close as possible to the device to continuously tune the polarization state from radial to azimuthal polarization while minimizing laser pointing drifts. Indeed, little laser drifts at the input can induce changes in the waveguide coupling conditions, specifically altering the propagating intensity. As HHG is an inherently highly nonlinear process, such variations would consequently lead to significant fluctuations in the EUV output.

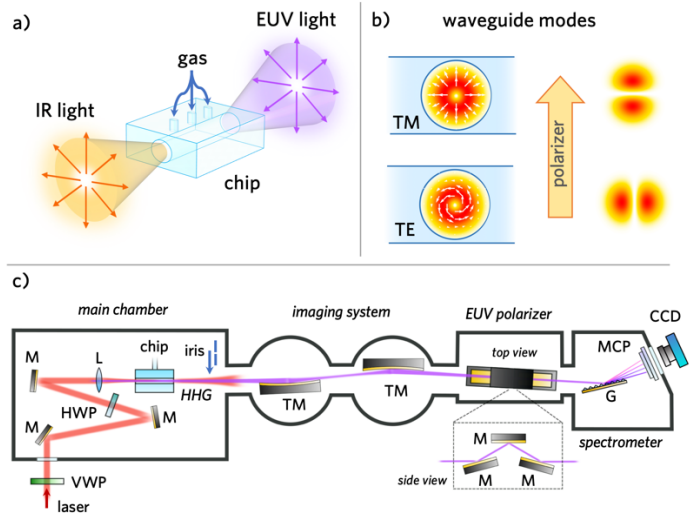

Figure 1. a) 3D illustration of the concept: the laser infrared (IR) vector beam is focused into the chip and via HHG it produces an EUV vector beam. This EUV beam maintains the same polarization state as it emerges at the exit. b) Visual representations of the $TE_{01}$ and $TM_{01}$ modes within the waveguide, along with the corresponding output intensity profiles after passing through a polarizer oriented in both instances as depicted in the figure. c) The complete experimental arrangement involves: the Ti:Sapphire laser entering the chamber through an anti-reflection-coated window, the vector waveplate (VWP), metallic mirrors (M), half-waveplate (HWP), $f = +200$ mm focusing lens (L), toroidal mirrors (TM), EUV concave grating (G), the micro-channel plate detector (MCP).

The last component is an anti-reflection coated lens with a focal length $f = +200$ mm that focuses the beam into the chip. The focusing lens was chosen according to calculations involving the overlap integral between the vector laser beam and the waveguide's $TM_{01}$ (or $TE_{01}$) modes (refer to Supplement 1, section 1), and allowed to achieve about 70% coupling efficiency. The microfluidic device, fabricated from silica through FLICE technique, consists of distinct components: a gas reservoir chamber, a central channel measuring 130 μm in diameter and 8 mm in length, and four microchannels connecting the reservoir to the central channel for the injection of argon gas at static pressure [32]. Located approximately 31 cm from the waveguide exit, an iris with a 1-mm aperture is mounted on a motorized stage to selectively transmit specific portions of the output beam. When the argon gas is injected the resulting EUV light is collected by an imaging system encompassing a pair of gold toroidal mirrors. This configuration arises from its primary utilization in transient spectroscopy experiments; however, it is not strictly indispensable for this investigation. An EUV polarizer made by three flat gold mirrors (see Supplement 1, section 3) favors the transmission of the horizontal polarization (with respect to the optical table) component of the EUV beam. Finally, the EUV light is diffracted by a concave gold-coated grating (*Hitachi*) onto a micro-channel plate detector equipped with a phosphorus screen. The resulting images are then captured using a CCD camera (Apogee Ascent A1050 – *Andor*). The latter three elements function as a spectrum analyzer operating within the 5 – 100 nm range (equivalent to 12 – 250 eV). The entire setup is maintained under vacuum conditions with a pressure ranging between $10^{-4}$ and $10^{-6}$ mbar.

Confirming the polarization state of the infrared vector beam is straightforward using a linear polarizer. In the case of radial polarization, a two-lobe intensity distribution aligned with the polarizer orientation is expected. Conversely, for azimuthal polarization, the two-lobe intensity distribution would manifest orthogonal to the polarizer direction (see Fig. 1b). To validate the effective guidance of $TM_{01}$ (or $TE_{01}$) modes within the channel, we employed a CMOS camera (beamage3.0 – *Gentec-EO*) to capture the intensity profiles at distinct locations, as depicted in Fig. 2. The resulting images clearly show the characteristic "donut" shapes associated with radial (or azimuthal) polarization. This is due to the presence of a polarization singularity at the center of the beam, compelling the electric field to attain zero intensity at that specific point. We configured the input polarization to be radial. To confirm the preservation of the laser's polarization state during propagation and to ensure that no uncontrolled mode mixing occurs within the device, we positioned an IR polarizer at the output. As depicted in Fig. 2, we observed a distinct two-lobe intensity distribution aligned with the polarizer's direction.

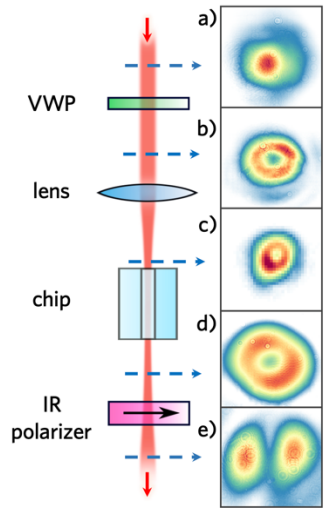

Figure 2. Intensity profiles of the IR laser beam measured at various positions (blue dotted arrows): a) before the vortex-waveplate, b) after the vortex-waveplate, c) focused at the entrance of the chip device, d) output of the chip (far-field), e) output far-field after an IR linear polarized.

Afterward, we removed the IR polarizer, leaving the incident polarization state radial. We set the maximum incident laser pulse energy to approximately 520 µJ, which is adequate to yield a good signal through HHG within the chip while still staying below the damage threshold of the vortex plate. Given a 70% coupling efficiency, the argon ionization potential of 15.7 eV and considering the specific energy distribution of the $TE_{01}$ (or $TM_{01}$) mode, we estimate, based on the semiclassical three-step model [35], a cut-off energy of about 56 eV (see Supplement 1, section 2).

As illustrated in Fig. 3a, we systematically adjusted the argon gas backing pressure across a range of 50 to 1000 mbar and recorded the corresponding harmonic intensity spectra. The peak of the HHG spectra exhibits a tendency to shift toward higher energy with increasing pressure and showcases a cut-off energy of approximately 50 eV, aligned with the theoretical expectations. The total EUV energy (normalized to its maximum) is determined by integrating the HHG spectra along the energy axis. This quantity, shown in Fig. 3b as a function of the gas pressure, reveals an increase in total yield until reaching approximately 300 mbar of gas pressure.

This behavior could be understood by considering the combined effect of light re-absorption and increased phase mismatch between the fundamental and harmonic fields with rising pressure, resulting in a reduction in the EUV yield [36]. Subsequently, having determined this pressure value as the optimum point, we systematically adjusted the laser pulse energy within the range of approximately 250 µJ to 520 µJ by means of a neutral density filter and recorded the corresponding HHG spectra. As illustrated in Fig. 3c, we observed a shift of the spectra peak towards higher energies accompanied by an extension of the cut-off energy, which exhibits a linear scaling as a function of laser pulse energy. The total EUV energy, shown in Fig. 3d, demonstrates an exponential growth with no signs of saturation. Hence, for subsequent experiments, we fixed the pulse energy at 520 µJ and the pressure at 300 mbar.

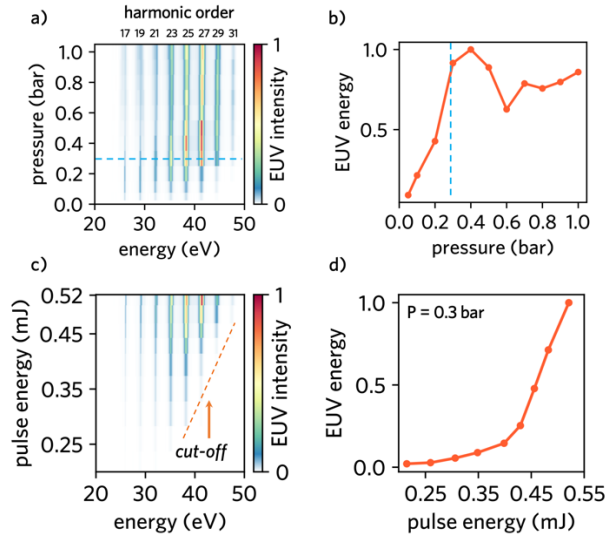

Figure 3. a) Normalized harmonics intensity as a function of the gas pressure. b) Normalized total harmonics energy versus gas pressure, the light blue dashed line represents the optimum point, corresponding to a pressure of 300 mbar. c) Normalized harmonics intensity as a function of the laser pulse energy at 300 mbar pressure. d) Normalized total harmonics as a function of the laser pulse energy.

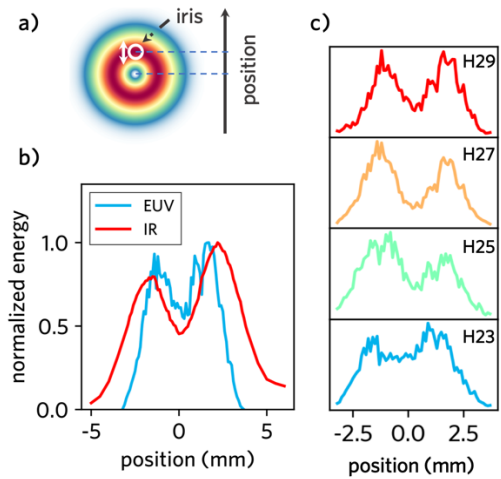

Figure 4. a) Illustration of the measurement performed. b) Normalized total transmitted energy of the fundamental laser beam at 800 nm (red) and EUV generated light via HHG (blue) as a function of the iris position. c) Normalized transmitted energy of some selected harmonics as a function of the iris position.

To confirm the vectorial nature of the output EUV beam, the initial step is to examine whether there is a singularity at the center. For this purpose, we moved the iris in the vertical direction while measuring in two separated sessions the transmitted IR and EUV energy, as illustrated in Fig. 4a. To measure the transmitted IR power, we positioned a mirror immediately after the iris. This was followed by a lens to precisely direct and focus the transmitted radiation onto a power meter situated inside the chamber at ambient pressure. In this setup, the power of the laser incident on the chip was deliberately reduced to mitigate any potential plasma-related effects within, and in close proximity of, the chip channel. For the EUV, we integrated the recorded HHG spectra. The outcomes of this measurement are illustrated in Fig. 4b. Notably, the EUV energy profile faithfully mirrors the IR counterpart, distinctly revealing a dip at the center. Figure 4c further displays the energy distributions for the four most intense harmonics. They exhibit analogous features in corresponding positions, indicating a similar divergence among all the harmonics. While the EUV energy profile with a dip in the center suggests the vector nature of the beam, it is not conclusive enough to unequivocally assert it.

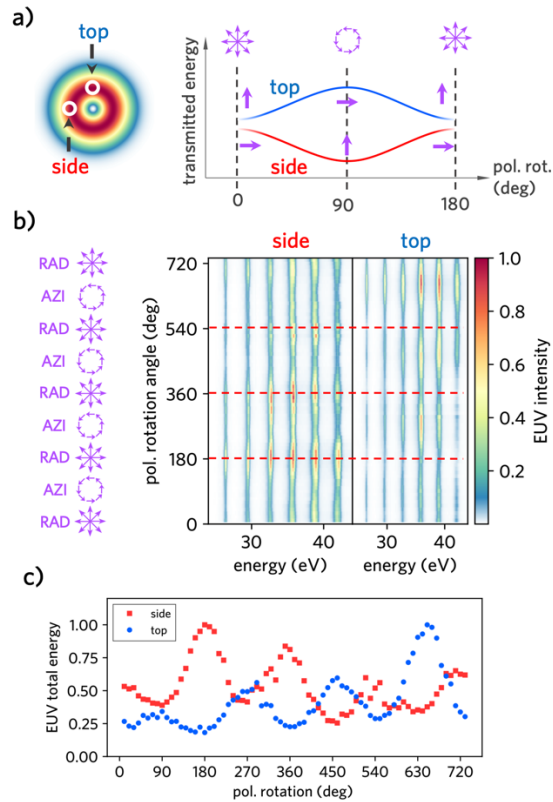

Figure 5. a) Visualization of the conducted experiment. b) 2D normalized intensity maps depicting EUV intensity spectra through an iris positioned atop and alongside the output EUV beam, varied with polarization rotation angle. c) Normalized total transmitted EUV energy plotted against the polarization rotation angle at the corresponding two locations along the output EUV beam.

To verify the polarization state of the generated EUV vector beam, we systematically recorded the HHG spectra by varying the HWP rotation angle, representing a continuous transition of the laser beam's polarization state from radial to azimuthal. The rotation was performed in increments of 5 degrees, corresponding to an effective 10-degree polarization rotation. This measurement was conducted with the iris positioned at two key points in the beam: the top and side (see Fig. 5a). Indeed, these two positions are anticipated to locally possess orthogonal polarization states. Considering that the EUV polarizer favors horizontal polarization (parallel to the optical table, i.e., the *s*-polarization – see Supplement 1, section 3) over vertical polarization, we expect to observe oscillations in harmonic intensity at these two locations as a function of the polarization rotation angle. Importantly, these oscillations should be precisely out of phase and exhibit a consistent 90-degree phase difference (see Fig. 5a).

In Fig. 5b, the HHG spectra are presented during a complete rotation of the HWP, corresponding to a polarization rotation of 720 degrees. Notably, the harmonic intensity exhibits clear oscillations with the rotation angle, where those transmitted to the iris atop the beam oscillate with a 90-degree phase shift compared to those sampled aside. Moving to Fig. 5c, the total energy of the EUV light is depicted, obtained by numerically integrating the transmitted HHG spectra at the same two locations. The two transmitted EUV energies oscillate out of phase as expected. The observed intensity fluctuations are mostly due to slight laser pointing drifts at the chip entrance induced during HWP rotation. This conclusive measurement unmistakably confirms the vectorial nature of the EUV beam, showcasing a continuous evolution of the output beam's polarization from radial to azimuthal.

## 3. Conclusion

We have successfully demonstrated a compact table-top source of coherent EUV vector beams generated inside a femtosecond-laser-micromachined microfluidic device. Initially, vector beams (e.g., with radial polarization) of ultrashort and intense IR pulses were generated through a liquid-crystal-based vortex plate. These beams were then focused into the chip, where we observed the effective decomposition of the laser's polarization state into a superposition of $TE_{01}$ and $TM_{01}$ modes within the waveguide. Through the HHG process, these modes were effectively upconverted into the EUV range, resulting in the synthesis of an equivalent vector beam at the output of the device. Our approach not only allows for the generation of EUV vector beams within a compact platform but also eliminates the need for bulky and inefficient solutions like standard gas jet nozzles. Furthermore, our chip has already demonstrated its capability to reach the soft X-ray regime, producing radiation extending up to 200 eV photon energy using helium gas [32]. Expanding the spectral range into the water window (approximately between 280 eV – 530 eV), essential for biological research, could be achieved by employing high-repetition-rate mid-infrared driving lasers, such as commercially available Yb-laser-pumped optical parametric chirped-pulse amplification systems. In contrast to standard approaches, our chip stands out as a unique platform for manipulating vector beam properties directly through the modal properties of the waveguide. This capability includes: i) controlling gas density distribution along the channel [31], ii) implementing quasi-phase-matching techniques through periodic channel modulation [37], iii) utilizing asymmetrical core geometry to affect differently the propagation constant of $TE_{01}$ and $TM_{01}$ modes, and iv) coupling into the chip more complex vector beams, possibly created with two or more wavelengths. Moreover, these devices enable the direct on-chip realization of numerous gas spectroscopy studies aligning with the paradigm of lab-on-chip technologies. These capabilities represent a significant advancement in the generation of customized and complex light fields from the EUV to soft X-rays as well as in the miniaturization and integration of attosecond spectroscopic experiments.

**Funding.**
This project has received funding from the European Union Horizon 2020 research and innovation program under grant agreement No 964588 (X-PIC), by the European Research Council MSCA-ITN SMART-X (Grant No. 860553), by the European Union's NextGenerationEU Programme with the I-PHOQS Infrastructure [IR0000016, ID D2B8D520, CUP B53C22001750006] "Integrated infrastructure initiative in Photonic and Quantum Sciences" and with the PRIN 2022 20224KAC28 "CHANGE" and the PRIN 2022 PNRR P20224AWLB "HAPPY".

**Disclosures.** The authors declare no conflicts of interest.

**Data availability.** The data that support the findings of this study are available from the corresponding author upon reasonable request.

**Supplemental document.** See Supplement 1 for supporting content.

# Synthesizing extreme-ultraviolet vector beams in a chip: supplemental document

**Riccardo Piccoli,**[1,2,*] **Marco Bardellini,**[1] **Stavroula Vovla,**[1] **Linda Oberti,**[1] **Kamal A. A. Abedin,**[1] **Anna G. Ciriolo,**[3] **Rebeca Martínez Vázquez,**[3] **Roberto Osellame,**[3] **Luca Poletto,**[4] **Fabio Frassetto,**[4] **Davide Faccialá,**[3] **Michele Devetta,**[3] **Caterina Vozzi,**[3] and **Salvatore Stagira**[1]

[1]*Department of Physics, Politecnico di Milano, Milano, I-20133, Italy*
[2]*Department of Molecular Sciences and Nanosystems, Ca' Foscari University of Venice, Venice, 30172 Italy*
[3]*National Research Council (CNR), Institute for Photonics and Nanotechnologies, Milano, I-20133, Italy*
[4]*National Research Council (CNR), Institute for Photonics and Nanotechnologies, Padua, I-35131 Italy*
*\*riccardo.piccoli@unive.it*

## 1. Coupling free-space vector beams into the chip waveguide

Without reporting a complete reiteration of the hollow-core waveguides theory [1], we recall that, under specific approximations, the electric field expressions of the $EH_{11}$, $TM_{01}$, and $TM_{01}$ and modes are shown in Fig. S1 together with their intensity profiles.

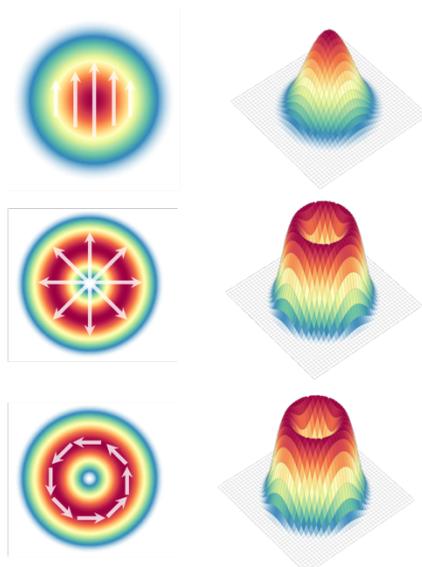

$$EH_{11}: \begin{cases} E_\theta = J_0\left(\frac{u_{11}r}{a}\right)\cos\theta \\ E_r = J_0\left(\frac{u_{11}r}{a}\right)\sin\theta \ ; \ u_{11} \cong 2.405 \\ E_z = -i\frac{u_{11}}{ka}J_1\left(\frac{u_{11}r}{a}\right) \end{cases}$$

$$TM_{01}: \begin{cases} E_\theta = 0 \\ E_r = J_1\left(\frac{u_{01}r}{a}\right) \ ; \ u_{01} \cong 3.832 \\ E_z = i\frac{u_{01}}{ka}J_0\left(\frac{u_{01}r}{a}\right) \end{cases}$$

$$TE_{01}: \begin{cases} E_\theta = J_1\left(\frac{u_{01}r}{a}\right) \\ E_r = 0 \ ; \ u_{01} \cong 3.832 \\ E_z = 0 \end{cases}$$

Figure S1. (Left). Expressions for the components of the electric fields associated with the modes. (Right) Top and 3D views illustrating the intensity profiles of the modes. The direction of the electric field is indicated by white arrows.

where $k = \frac{2\pi}{\lambda}$ is the wavevector of light in vacuum, $a$ denotes the waveguide radius, and $J_x$ the Bessel function of the first kind.

The coupling efficiency $\eta$ of the chip can be estimated by calculating the mode overlap between the experimentally recorded image of the laser vector beam intensity profile at the focus $|E_{VB}|^2$ (immediately in front of the entrance of the chip, at ambient pressure and attenuated laser intensity) and the theoretically computed TM$_{01}$ mode intensity distribution $|E_{TM}|^2$ (see Fig. S2) as follows:

$$\eta = \frac{|\int E_{VB}^* E_{TM}\, dA|^2}{\int |E_{VB}|^2 dA \int |E_{TM}|^2 dA} \cong 85\%$$

The theoretically predicted value aligns closely with the experimentally measured overall transmission of the laser light through the chip, which stands at approximately 70%.

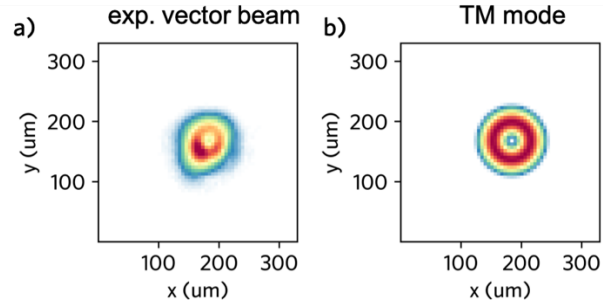

Figure S2. a). Experimentally recorded intensity profile of the laser vector beam at the chip entrance focus. b) Calculated intensity distribution of the TM$_{01}$ (or TE$_{01}$) mode in the waveguide.

## 2. Peak intensity of the EH$_{11}$, TE$_{01}$, and TM$_{01}$ modes

Assuming the profiles in the Fig. S1 depict the fluence, denoted as $F(r)$, of the beam in terms of irradiance (J/m²) as a function of the radial coordinate $r$, the total energy (in joules) carried by the beam can be derived as follows:

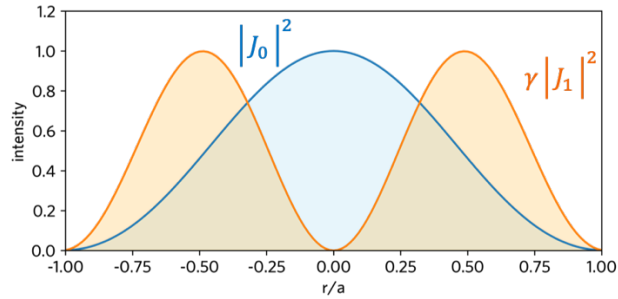

Figure S3. Radial cross-sections depicting the normalized intensity profiles of EH$_{11}$ mode (blue) and TM$_{01}$ (or TE$_{01}$) mode (orange).

$$\mathcal{E} = \int_0^a F(r) 2\pi r\, dr$$

Therefore:

$$\mathcal{E}_{EH} = \int_0^a \left|J_0\left(r\frac{u_{11}}{a}\right)\right|^2 2\pi r\, dr = \pi a^2 J_1(u_{11})^2 \cong 0.85 a^2$$

$$\mathcal{E}_{TM} = \gamma \int_0^a \left|J_1\left(r\frac{u_{01}}{a}\right)\right|^2 2\pi r\, dr = -\gamma \pi a^2 J_0(u_{01}) J_2(u_{01}) \cong 1.5 a^2$$

To facilitate a comparison of the energy contained in both modes, it is beneficial to standardize the peak intensity of both modes to 1, as illustrated in Fig. S3. Given that HHG is contingent upon peak intensity, it is advantageous to assess the two modes under identical peak intensity conditions. The maximum of mode EH$_{11}$, as described by the function $J_0$, is already set at 1. However, the squared modulus of the mode TM$_{01}$, represented by $J_1$, is not. Thus, we can scale the squared modulus by a normalization parameter $\gamma \cong 2.95$ to achieve a peak fluence $F_{pk} = \max(\gamma |J_1(x)|^2) = 1$ (J/m$^2$). The ratio $\xi$ between the two energies is then expressed as:

$$\xi = \frac{\mathcal{E}_{TM}}{\mathcal{E}_{EH}} = -\frac{\gamma J_0(u_{01}) J_2(u_{01})}{J_1(u_{11})^2} \cong 1.77$$

Hence, when considering equivalent energy, the peak fluence of the radially polarized TM$_{01}$ mode (or azimuthally polarized TE$_{01}$ mode) is 1.78 times lower compared to the linearly polarized EH$_{11}$ mode. Given the peak fluence, $F_{pk}$, for a Gaussian pulse envelope in time, the peak intensity is determined as follows:

$$I_{pk} = \frac{2 F_{pk}}{t_p} \sqrt{\frac{\ln 2}{\pi}}$$

where $t_p$ denotes the full-width at half-maximum (FWHM) of the pulse duration. The relationship between the electric field and the peak intensity is expressed as:

$$E_{pk}^2 = \frac{2 I_{pk}}{c \varepsilon_0}$$

where $c$ represents the speed of light, and $\varepsilon_0$ denotes the permittivity of vacuum. The cut-off energy for the EUV light through HHG is calculated following the three-step model [2] as:

$$h\nu_{cut-off} \cong I_p + 3.17 U_p \qquad U_p = \frac{e^2 E_{pk}^2}{4 m_e \omega_0^2}$$

where $I_p$ is the ionization potential of the gas filling the channel, $U_p$ is the ponderomotive energy, $e$ and $m_e$ denote the electron charge and mass, respectively, $h$ stands for the Planck constant, $E_{pk}$ corresponds to the peak electric field of the laser field envelope, and $\omega_0$ represents the laser angular frequency. To be specific, utilizing the parameters $I_p = 15.7$ eV (argon), $\omega_0 = 2.35 \cdot 10^{15}$ rad/s (corresponding to a laser wavelength of 800 nm), $t_p = 25$ fs, $\mathcal{E}_{TM} = 520 \cdot 0.7 = 364$ µJ (where 0.7 accounts for the experimentally measured coupling efficiency), $I_{pk} = 2.15 \cdot 10^{14}$ W/cm$^2$, we get a $h\nu_{cut-off} \cong 56$ eV. A final observation pertains to the presence of the z-component of the electric field in the TM mode (and EH). In principle, this component has the potential to deactivate the HHG process, as observed in circular polarization. However, given that the waveguide radius significantly exceeds the laser wavelength (i.e., $ka \gg 1$), this component become very small and, as such, can be considered negligible.

## 3. Polarization contrast of the EUV polarizer

In order to establish a clear polarization contrast between *p*- and *s*-polarization, a crucial step in validating the vectorial nature of the generated EUV beam, we built a polarizer using three gold mirrors [3] configured as illustrated in Fig. S4.

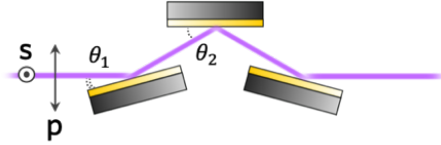

Figure S4. Polarizer geometry (side view).

The field reflection coefficients $r$ can be derived from the Fresnel's equations:

$$r_s = \frac{n_2 \cos\theta_i - n_1 \cos\theta_t}{n_2 \cos\theta_i + n_1 \cos\theta_t} \qquad r_p = \frac{n_2 \cos\theta_t - n_1 \cos\theta_i}{n_2 \cos\theta_t + n_1 \cos\theta_i}$$

The incident and transmitted angles are connected through the Snell's law: $n_1 \sin\theta_i = n_2 \sin\theta_t$. For our calculations, we extracted the complex refractive index of gold, $n_2$, within the energy range of 30 eV to 50 eV (see Fig. S5) from the freely accessible CXRO database online [4] while $n_1$ is set to 1. We considered a grazing incident angle of $\theta_1 = 10°$ and, for geometric considerations, set $\theta_2 = 2\theta_1$. We then computed the overall power transmission (accounting for three reflections) of the polarizer, denoted as $T$, as a function of the input polarization rotation angle $\phi$:

$$T(\phi) = |r_{p,t} \cos\phi|^2 + |r_{s,t} \sin\phi|^2$$

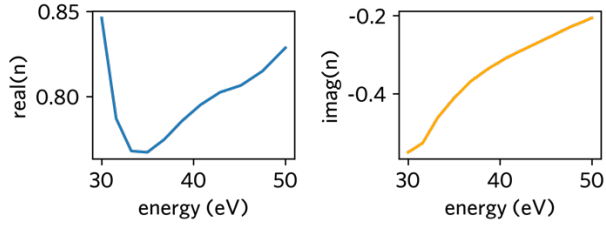

Figure S5. Real and imaginary parts of the complex refractive index of gold.

Where $r_{p,t} = r_{p,1} r_{p,2} r_{p,3}$ represents the product of the three reflection coefficients corresponding to the three mirrors. Figure S6 illustrates the outcomes of this calculation: a contrast of approximately 2.5 can be achieved between *s*- and *p*-polarization.

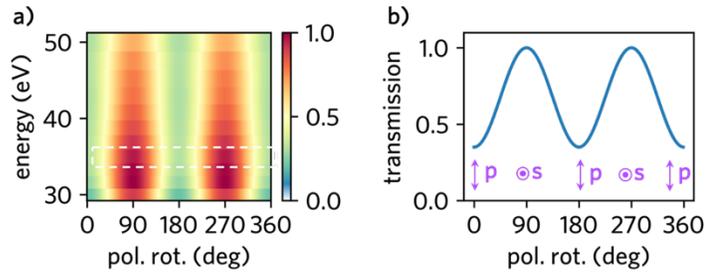

Figure S6. a) Calculated and normalized polarizer transmission versus polarization angle. b) Transmission at approximately 35 eV extracted from the region delineated by the dashed white rectangle in (a).

To validate the theoretical prediction, we generated XUV through HHG without the vortex waveplate, i.e., utilizing linearly polarized light. We recorded HHG spectra while varying the rotation of the half-wave plate (i.e., adjusting the incident polarization of the generated XUV light on the polarizer). The results are presented in Fig. S7 for a complete 360-degree rotation. Notably, the intensity of the harmonics clearly oscillates as the polarization rotates. A contrast of approximately 2.5 between *s*- and *p*-polarization is evident, in good agreement with the theoretical predictions.

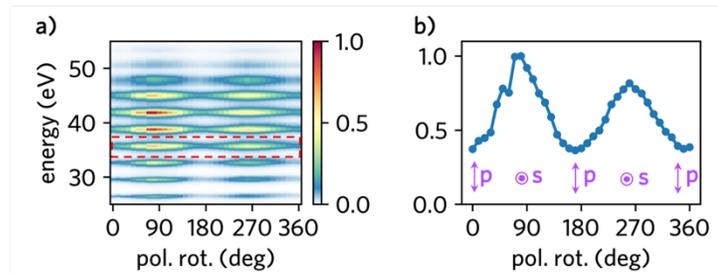

Figure S7. a) Transmitted experimental HHG intensity spectra versus XUV polarization angle. b) Transmission at approximately 35 eV extracted from the region delineated by the dashed red rectangle in (a).